%% file: ICSE2023.tex
\documentclass[10pt,conference]{IEEEtran}
\usepackage{amsmath,amssymb,amsfonts}
\usepackage{graphicx}
\usepackage{textcomp}
\usepackage{xcolor}
\usepackage{comment}
\usepackage{multirow}
\usepackage{hyperref}
\hypersetup{
    colorlinks=true,
    linkcolor=black,
    citecolor=black,
    urlcolor=blue
    }
\def\BibTeX{{\rm B\kern-.05em{\sc i\kern-.025em b}\kern-.08em
    T\kern-.1667em\lower.7ex\hbox{E}\kern-.125emX}}
    
\newcommand{\tabincell}[2]{\begin{tabular}{@{}#1@{}}#2\end{tabular}}

\makeatletter
\newcommand{\linebreakand}{%
  \end{@IEEEauthorhalign}
  \hfill\mbox{}\par
  \mbox{}\hfill\begin{@IEEEauthorhalign}
}
\makeatother

\begin{document}

\title{SeeHow: Workflow Extraction from Programming Screencasts through Action-Aware Video Analytics}

\author{\IEEEauthorblockN{1\textsuperscript{st} Dehai Zhao}
\IEEEauthorblockA{\textit{Australian National University} \\
\textit{and CSIRO's data61}\\
Australia \\
dehai.zhao@anu.edu.au}
\and
\IEEEauthorblockN{2\textsuperscript{nd} Zhenchang Xing}
\IEEEauthorblockA{\textit{Australian National University} \\
\textit{and CSIRO's data61}\\
Australia \\
zhenchang.xing@data61.csiro.au}
\and
\IEEEauthorblockN{3\textsuperscript{rd} Xin Xia}
\IEEEauthorblockA{\textit{Software Engineering Application} \\
\textit{Technology Lab, Huawei}\\
China \\
xin.xia@acm.org}
\linebreakand
\IEEEauthorblockN{4\textsuperscript{th} Deheng Ye}
\IEEEauthorblockA{\textit{Tencent AI Lab} \\
China \\
dericye@tencent.com}
\and
\IEEEauthorblockN{5\textsuperscript{th} Xiwei Xu}
\IEEEauthorblockA{\textit{CSIRO's data61} \\
Australia \\
xiwei.xu@data61.csiro.au}
\and
\IEEEauthorblockN{6\textsuperscript{th} Liming Zhu}
\IEEEauthorblockA{\textit{CSIRO's data61} \\
Australia \\
liming.zhu@data61.csiro.au}
}

\maketitle

\begin{abstract}
Programming screencasts (e.g., video tutorials on Youtube or live coding stream on Twitch) are important knowledge source for developers to learn programming knowledge, especially the workflow of completing a programming task. 
Nonetheless, the image nature of programming screencasts limits the accessibility of screencast content and the workflow embedded in it, resulting in a gap to access and interact with the content and workflow in programming screencasts. 
Existing non-intrusive methods are limited to extract either primitive human-computer interaction (HCI) actions or coarse-grained video fragments.
In this work, we leverage Computer Vision (CV) techniques to build a programming screencast analysis tool which can automatically extract code-line editing steps (enter text, delete text, edit text and select text) from screencasts.
Given a programming screencast, our approach outputs a sequence of coding steps and code snippets involved in each step, which we refer to as programming workflow. 
The proposed method is evaluated on 41 hours of tutorial videos and live coding screencasts with diverse programming environments.
The results demonstrate our tool can extract code-line editing steps accurately and the extracted workflow steps can be intuitively understood by developers.
\end{abstract}


\input{Introduction}

\input{Approach}

\input{Evaluation}
\input{RelatedWork}

\bibliographystyle{IEEEtran}
\bibliography{reference}

\end{document}

%% file: Introduction.tex
\section{Introduction}
\label{sec:introduction}

\begin{figure}
	\centering
	\includegraphics[width=\linewidth]{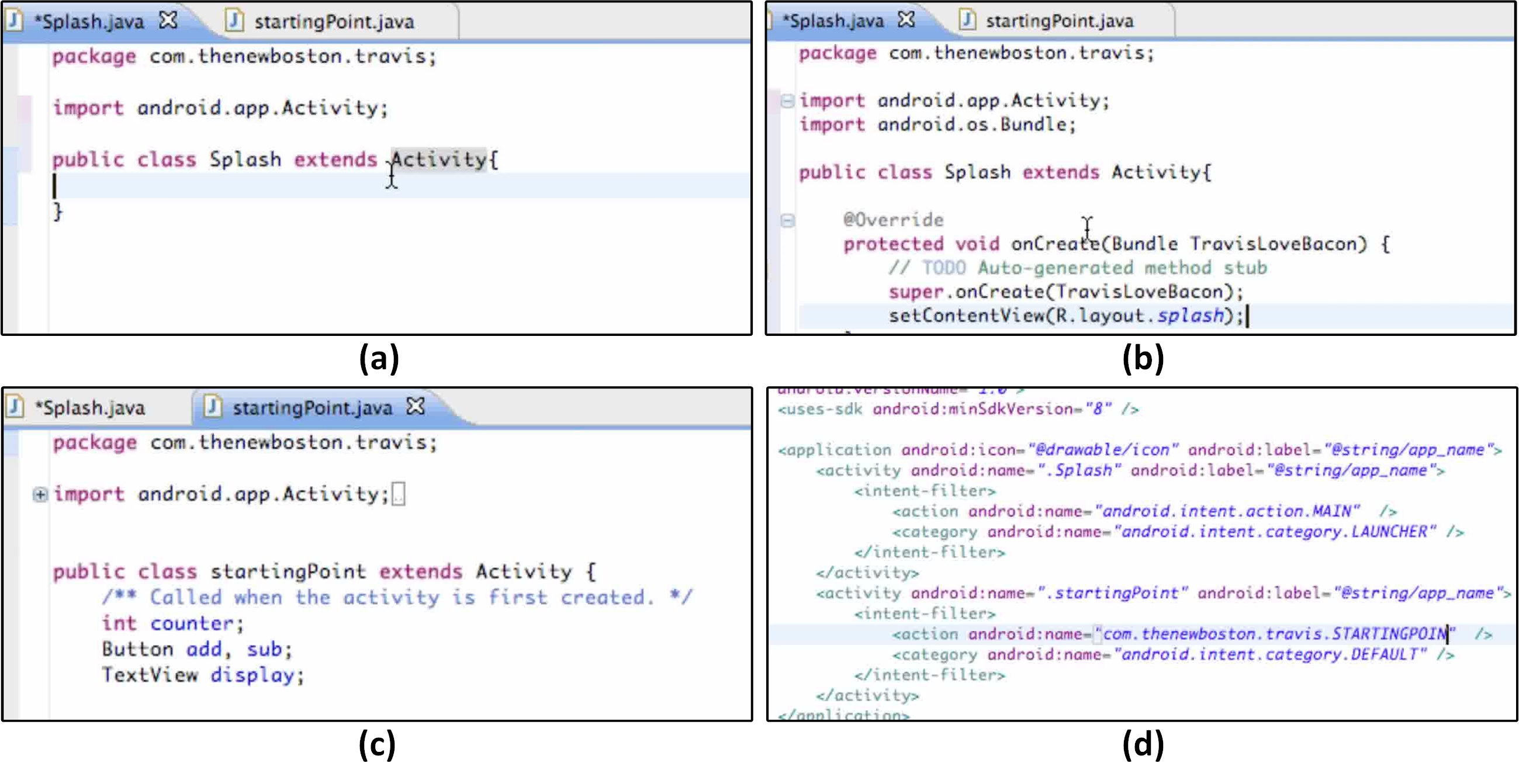}
	\vspace{-2em}
	\caption{An Example of Programming Workflow}
	\vspace{-1.5em}
	\label{fig:workflowexample}
\end{figure}

Computer programming involves two types of knowledge: \textit{knowing-what} (a.k.a declarative knowledge) and \textit{knowing-how} (a.k.a procedural knowledge).
Fig.~\ref{fig:workflowexample} illustrates an example.
Knowing-what involves facts or propositions of specific programming concepts or entities.
For example, \textsf{Activity} in Fig.~\ref{fig:workflowexample}(a) is an Android class which takes care of creating a window in which developers can place app UI, and \textsf{Bundle} in Fig.~\ref{fig:workflowexample}(b) is a mapping from string keys to various parcelable values.
Knowing-how involves the workflow to complete a programming task step by step.
For example, the developer first creates a class by entering the code line ``public class Splash extends Activity'' (Fig.~\ref{fig:workflowexample}(a))\footnote{The video is available at \href{http://seecollections.com/seehow/example}{http://seecollections.com/seehow/example}}, and then enter the code block ``public void onCreate(...) \{super.onCreate();\}'' to override \textsf{onCreate()} and enter one line of API call ``setContentView(R.layout.Splash)'' (Fig.~\ref{fig:workflowexample}(b)).
Next, she creates another activity class \textsf{StartingPoint} and declares several fields (Fig.~\ref{fig:workflowexample}(c)) by entering several code lines.
Finally, she sets the proper bundling mapping by selecting the text ``MAIN'' in the bundle mapping XML file, and then editing it to ``STARTINGPOINT'' (Fig.~\ref{fig:workflowexample}(d)).
In this work, we refer to such code-line editing as coding step (or workflow step interchangeably).
The editing of several code lines (a code block) at the same time (e.g., delete or paste several code lines) is considered as one code-line editing step.

Programming screencasts are screen recordings of the developers performing programming tasks.
Typical examples include programming videos on YouTube and live coding streams on Twitch.
Programming screencast is an important medium for learning programming know-what and know-how knowledge.
For example, the top 1 Python tutorial on Youtube has been watched for 28,802,076 times.
The workflow (coding steps, best practices, emergent errors and fixes) in programming screencasts complement know-what programming knowledge (e.g., API documentation, Stack Overflow discussions) with know-hows~\cite{ponzanelli2016too, bao2018vt}.
The know-how knowledge is also important for task-oriented knowledge recommendation~\cite{sun2019ai, sun2020task}

In spite of its importance, \textit{the workflow is implicitly embedded in the screencasts}, which greatly limits the access and interaction with programming workflow in programming screencasts.
This makes it hard for developers to get a quick overview of the workflow, or to navigate and search the workflow steps~\cite{bao2015activityspace,bao2018vt}.
This also makes it impossible to support advanced workflow analysis.
For example, by comparing wrong and correct workflow, the developer may find that she does not set activity bundle after creating a new activity class, which cause the failure of starting the activity.
To support more effective use of programming screencasts in software engineering tasks, several efforts have been made~\cite{ponzanelli2016too, khandwala2018codemotion, zhao2019actionnet, bao2015activityspace,bao2018vt}.
These methods make the workflow in programming screencasts explicit by software instrumentation or computer vision techniques. 

First, coding steps can be recorded by instrumenting software tools or operating systems~\cite{bao2018vt}.
Such instrumentation-based methods are intrusive and constrained by the availability of accessibility APIs~\cite{bao2015scvripper}.
Second, computer vision techniques have been used to extract or infer programming actions~\cite{ponzanelli2016too, khandwala2018codemotion, zhao2019actionnet}.
Such techniques are non-intrusive, but they extract either primitive HCI actions between two consecutive frames~\cite{zhao2019actionnet} or coarse-grained activities involving similar contents~\cite{ponzanelli2016too, khandwala2018codemotion}. 
Primitive HCI actions are too fine-grained to correspond to the intuitive understanding of coding steps, and content-similarity based video fragments are too coarse-grained and one such fragment will involve many code-line changes.
In a comparative study of 135 developers, Bao et al.~\cite{bao2018vt} show that coding steps at line granularity are more effective for learning know-how knowledge than coarse-grained programming activities.
Unfortunately, coding steps at line granularity is currently supported by only instrumentation-based method.

To fill the gap in between the existing primitive and coarse-grained non-intrusive workflow extraction methods, we propose a novel action-aware workflow extraction approach that extracts code-line editing steps from programming screencasts.
In particular, our approach extracts four types of coding steps: \textit{enter text}, \textit{delete text},  \textit{edit text}, and \textit{select text}.
Text can be source code or other textual content (e.g., command-line command, xml content, and text field input).
We develop a deep-learning based computer vision method to recognize and aggregate primitive HCI actions and corresponding text edits into coding steps at line granularity.


\begin{figure}
	\centering
	\includegraphics[width=\linewidth]{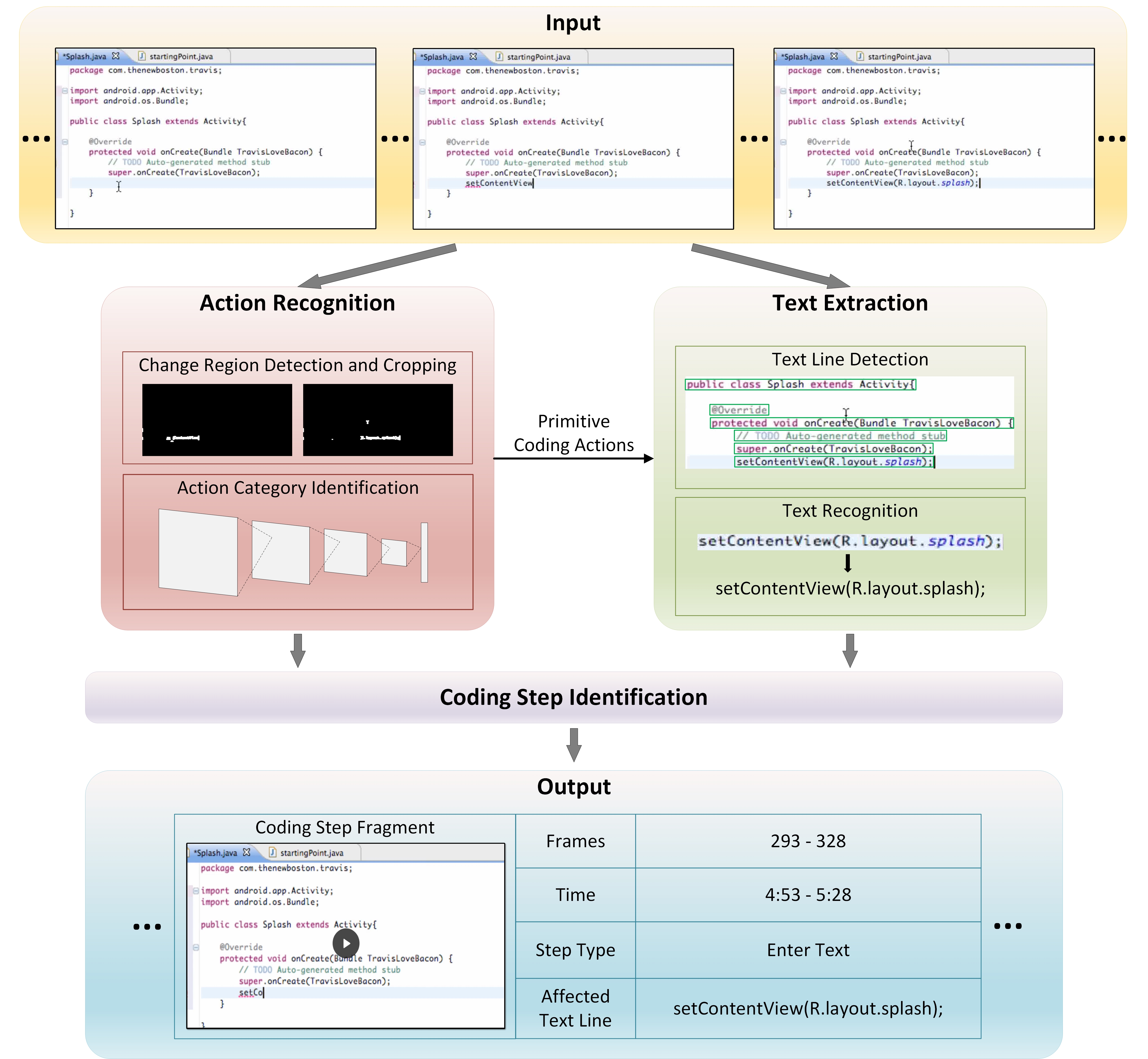}
	\vspace{-1.5em}
	\caption{Main Steps of Our Approach}
	\vspace{-1em}
	\label{fig:approach}
\end{figure}

To evaluate our method, we build a dataset of programming screencasts, including 260 programming videos from YouTube and 10 live coding streams from Twitch (see Table~\ref{tab:dataset}).
These screencasts involve multiple programming languages (e.g., Python and Java) and many different programming tasks (e.g., programming basics, Android app, web app, and game).
They were created by 8 different developers, using very different development tools (e.g., Eclipse, Pycharm, Sublime).
The total duration of these screencasts is 41 hours.
Two authors invest significant manual efforts to label the coding steps in these screencasts, including the start and end frame of each coding step and the text edited by the step\footnote{The tool and dataset are available at the GitHub repository \href{https://github.com/DehaiZhao/SeeHow}{https://github.com/DehaiZhao/SeeHow}}.
We obtain 5,466 coding steps with variant durations (7.62$\pm$10.82 seconds).

On this dataset, our workflow extraction method achieves 0.501 in F1-score at IoU = 1.
IoU measures the intersection over union between the video fragments of the identified steps and the ground-truth steps.
IoU = 1 means the perfect match of the two fragments.
For the not-perfectly-aligned identified steps, 94.2\% have only 0 or 1 frame offset at the start or the end.
Our evaluation also confirms the stability of our method for different programming languages, developers, and development tools.
We invite 6 developers from software industry to evaluate the quality of 2,605 not-perfectly-aligned coding steps.
They rate 83\% of these coding steps as correct, and their ratings have substantial agreements with Fleiss' kappa of 0.83. 
This suggests that the small misalignments of most of the not-perfectly-aligned steps do not affect the correct understanding of the identified steps.
At the end of the user study, we discuss with the developers about the potential scenarios to use our tool, such as software asset management and workflow logging .

Our work makes the following contributions:
\begin{itemize}
	\item We develop the first computer-vision based action-aware method for extracting coding steps at the line granularity from programming screencasts.
	
	\item We contribute a large dataset of programming screencasts with 5,466 manually labeled coding steps for evaluating computer-vision based workflow extraction methods.
	
	\item We conduct extensive experiments to evaluate the accuracy and generality of our method and the quality of the extracted coding steps for human understanding.
	
\end{itemize}

%% file: Approach.tex
\section{Approach}
\label{sec:approach}

Fig.~\ref{fig:approach} presents the main steps of our approach.
Our approach takes as input a programming screencast (i.e., a sequence of screenshots), and outputs a sequence of coding steps (Section~\ref{sec:input}).
We adopt computer-vision based techniques for action recognition and text extraction.
By analyzing the consecutive frames, action recognition extracts screenshot regions affected by primitive HCI actions and determines action categories (Section~\ref{sec:actionrecognition}), and text extraction extracts and aligns text lines affected by HCI actions (Section~\ref{sec:textextraction}).
The action and text information are then aggregated to determine coding steps and the corresponding video fragments (Section~\ref{sec:codingstepidentification}).


\subsection{Definition of Input and Output}
\label{sec:input}

An input programming screencast is a sequence of screenshots taken at a specific time interval (e.g., 1 second).
Each screenshot is a frame in the screencast.
Recording a screencast can be achieved by using operating-system level APIs, without the need for application-specific support.
Taking advantage of this generality, our approach does not constrain the types of programming tasks, nor the development tools and programming languages used in the programming tasks.
It also does not make any assumption about computer settings (e.g., screen resolution, window size and arrangement) and screenshot-taking interval.
Furthermore, our approach does not rely on any HCI information recorded by instrumentation-based methods (e.g., ActivitySpace~\cite{bao2015activityspace}).

\begin{figure}
	\centering
	\vspace{-1em}
	\includegraphics[width=\linewidth]{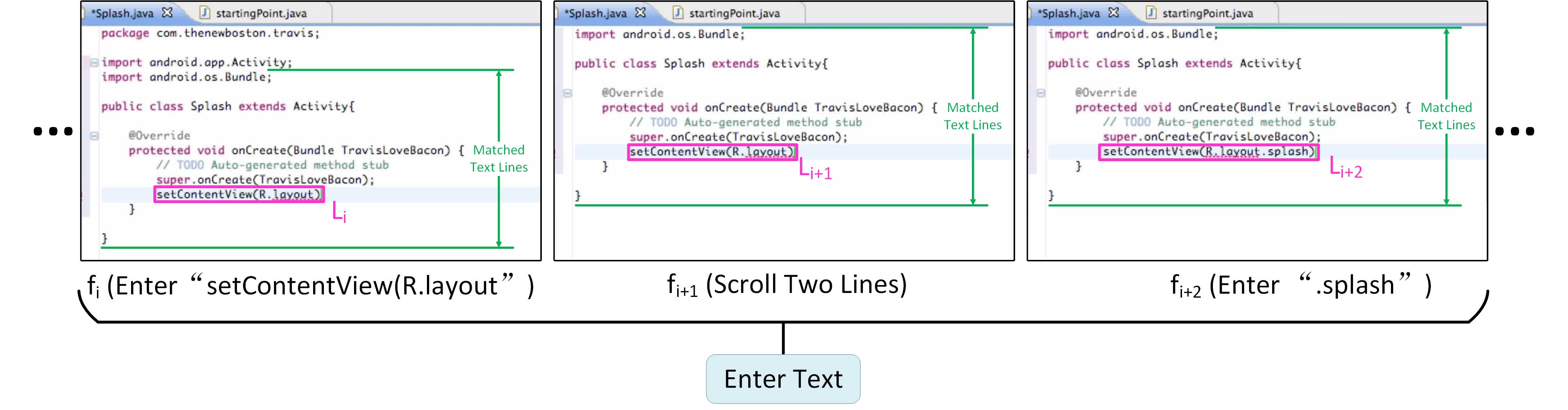}
	\caption{Example: scroll-content in between coding actions}
	\vspace{-1.5em}
	\label{fig:scroll}
\end{figure}

An output coding step consists of five pieces of information: the start and end frame, the start and end time, the corresponding video fragment, the type of coding step, and the text line affected by the coding step.
A sequence of identified coding steps constitutes a trailer of the programming screencast, illustrating its programming workflow.
We consider four types of coding steps: \textit{enter text}, \textit{delete text}, \textit{edit text} and \textit{select text}.
The text includes not only source code but also other software text, such as command-line commands, console output, text field input like file names or search keywords, XML file content, and web page content.
The granularity of text being manipulated is one line of text.
This line granularity represents basic and coherent steps in programming tasks, which is the granularity used for code versioning and patching.
If a block (several lines) of text is manipulated as an atomic unit we consider this block of text as a line of text.
Typical examples include a block of text being cut, pasted or selected as a whole, or a code block being added by code auto-completion or quick-fix assistants.   

\subsection{Action Recognition}
\label{sec:actionrecognition}

When identifying coding steps, our approach is aware of primitive HCI actions that constitute the coding steps.
This action awareness differentiates our approach from existing action-agnostic, content-based workflow extraction methods~\cite{ponzanelli2016codetube, khandwala2018codemotion, bao2020psc2code}.
It allows our approach to filter out irrelevant content changes on the computer screen, resulting from non-coding actions such as switch windows, trigger or leave pop-ups, which always confuse action-agnostic methods.
It also allows our approach to aggregate continual HCI actions on the same text line into code-line editing steps (see Section~\ref{sec:codingstepidentification}), which has never been achieved by content-only analysis.

In this work, we adopt ActionNet~\cite{zhao2019actionnet} to recognize primitive HCI actions in programming screencasts.
ActionNet recognizes three categories, nine types of frequent HCI actions in programming work:
\textit{move cursor/mouse} (move cursor, move mouse over editable text, and move mouse over non-editable text), \textit{edit chars} (enter chars and delete chars), \textit{interact with app} (select chars, scroll content, trigger or leave pop-ups, and switch windows).
All other HCI actions (e.g., move/resize windows, zoom-in/out text, and click UI elements like menu or list items) are recognized as an other-action type.

The consecutive frames capturing no actions, move-mouse or move-cursor are discarded, because they have no content changes.
For coding step analysis, enter/delete/select-chars are considered as coding-related actions.
Select-chars is also considered as coding related because the selected content is often edited afterwards.
Scroll-content, trigger/leave pop-ups, switch-windows and other-action are considered as non-coding-related actions, and the corresponding frames are discarded.

\subsection{Text Extraction} 
\label{sec:textextraction}

\begin{figure}
	\centering
	\vspace{-1em}
	\includegraphics[width=\linewidth]{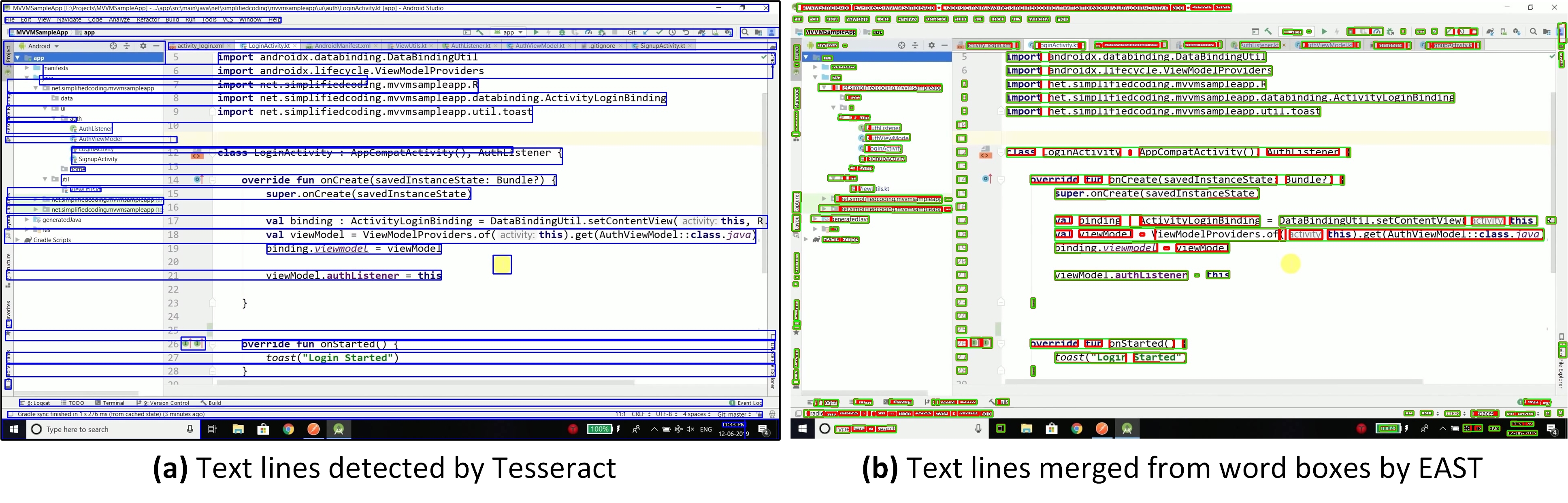}
	\caption{Example of text line detection}
	\vspace{-1.5em}
	\label{fig:textbox}
\end{figure}

If the two frames capture a coding-related action, our approach extracts and aligns text lines in the two frames.
We also extract and align text lines in the two frames capturing a scroll-content action.
Text-line alignment helps to aggregate continual primitive coding-related actions on the same text line, even they may be separated by small content scrolling in between (see Fig.~\ref{fig:scroll}).
Non-coding-related actions may change the displayed content (usually significantly).
But such screen changes are due to the appearance of new content or the update of the displayed content, rather than the actual content editing.
Our action-aware approach ignores such meaningless content changes resulting from non-coding actions.

\begin{figure*}
	\centering
	\includegraphics[width=\textwidth]{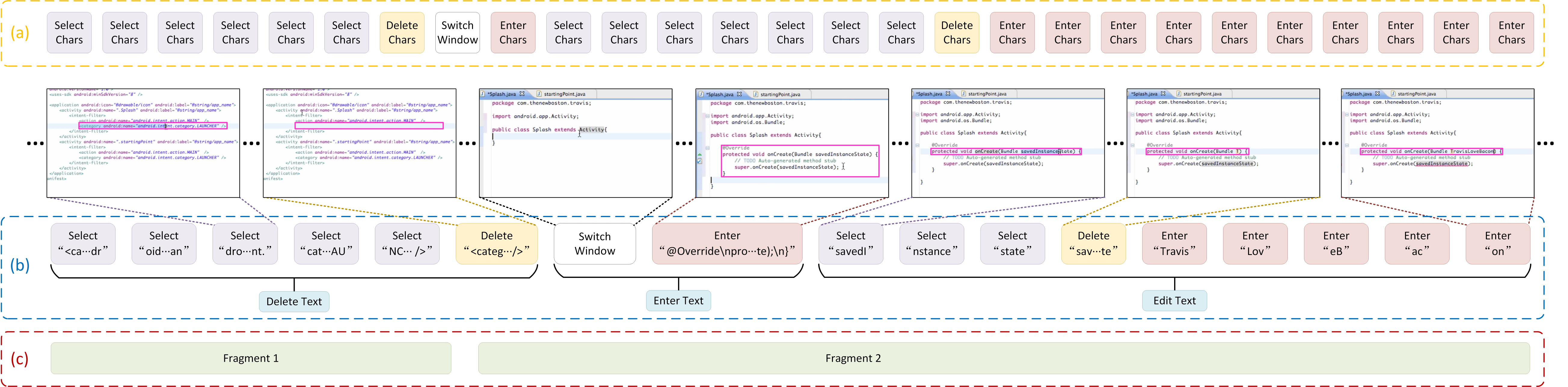}
	\caption{Illustration of three coding step identification methods. (a) ActionNet (primitive HCI actions). (b) SeeHow (code-line editing steps). (c) Output of CodeTube (coarse-grained video fragments without action labels)}
	\vspace{-1em}
	\label{fig:fragment}
\end{figure*}

Extracting text from UI screenshots has two sub-tasks: \textit{text line detection} - locate the bounding box of text elements (e.g., words) in the image, and \textit{text recognition} - convert a text image into text characters.
As shown in Fig.~\ref{fig:textbox}(a), UIs are usually composed of multiple windows.
Different windows use different text styles (e.g., font, size, alignment, color and contrast).
Even the content in the same window may use different text styles, for example, programming language keywords, constants or comments in code editor.
UIs also contain visual components such as menu, toolbar and tab which also display text.
All these UI text characteristics affect the choice of appropriate text extraction techniques.

Optical Character Recognition (OCR) tools (e.g., Tesseract~\cite{tesseract}) are commonly used for extracting text from UI screenshots~\cite{ponzanelli2016codetube, khandwala2018codemotion, bao2020psc2code}.
OCR tools are ideal for processing document images, but their performance degrades for UI screenshots~\cite{bao2020psc2code} due to complex view layouts and text styles.
Fig.~\ref{fig:textbox}(a) shows some inaccurate text lines on an IDE screenshot detected by Tesseract.
A recent large-scale empirical study on GUI widget detection~\cite{chen2020object} shows that UI text should be treated as scene text (usually in low resolution and with messy text layouts and styles) rather than document text.
Therefore, we adopt the state-of-the-art scene text detection tool EAST~\cite{zhou2017east} to detect text lines in UI screenshots.

EAST detects text at the word level, for example, the red word boxes in Fig.~\ref{fig:textbox}(b).
Word-level detection helps to distinguish component labels (e.g., menu text) from regular text (e.g., code).
Our approach scans the detected word text boxes from left to right and top to bottom, and merges the horizontally adjacent boxes iteratively.
Let $B_1(x_1, y_1, x_2, y_2)$ and $B_2(x_1, y_1, x_2, y_2)$ be the two adjacent text boxes, where $(x_1, y_1)$ is the top-left coordinate and $(x_2, y_2)$ is the bottom-right coordinate, and $B_2$ is right to $B_1$ (i.e., $B_2.x_1 > B_1.x_2$).
$B_1$ and $B_2$ will be merged into one text box if they satisfy the two conditions:
1) The vertical coordinate of the horizontal middle line of $B_2$ (i.e., $(B_2.y_1 + B_2.y_2) / 2$)  is within the top and bottom line of $B_1$ (i.e., $[B_1.y_1, B_1.y_2]$;
2) The minimum horizontal distance between $B_1$ and $B_2$ is less than the average width of the characters in the two boxes.
The average character width is computed as $((B_2.x_2 - B_2.x_1)+(B_1.x_2 - B_1.x_1))/num_c$ where $num_c$ is the total number of characters in $B_1$ and $B_2$ recognized by CRNN~\cite{shi2016end}.
If $B_2$ cannot be merged with $B_1$, $B_2$ is used as the new starting box to merge the word boxes to the right of $B_2$.
This merging process continues until no more word boxes are left to merge.
It outputs text lines such as those in green boxes in Fig.~\ref{fig:textbox}(b).

 

Given a detected text line, our approach crops the image region by the text-line bounding box and uses the CRNN tool~\cite{shi2016end} to convert the cropped text image into text characters.
CRNN stands for Convolutional Recurrent Neural Network, which is the state-of-the-art model for text recognition.
CRNN uses convolutional neural network to extract image features, which makes it robust to text images with various fonts, colors and backgrounds.

\subsection{Coding Step Identification}
\label{sec:codingstepidentification}

A code-line editing step may consist of some primitive HCI actions.
For example, entering a line of code may consist of a sequence of enter-chars actions, some delete-chars actions to correct typing errors, and also non-coding-related actions, such as move cursor, scroll content and pop-up interactions.
Based on the recognized primitive actions and the extracted text lines, our approach aggregates continual coding-related actions into code-line editing steps.

Fig.~\ref{fig:fragment} (b) illustrates the examples of coding step identification by our method.
Our approach scans the sequence of $N$ screenshots from beginning to end.
Recall that the frames capturing no actions or move-mouse/cursor actions have already been discarded.
Let $f_i$ be the current frame pointer ($1 \leq i \leq N-1$).
Let $act_i$ be the $i-th$ action in the sequence, captured in $f_{i}$ and $f_{i+1}$.
If $act_i$ is a non-coding-related action (e.g, the 3rd frame in Fig.~\ref{fig:fragment}), the approach increments the frame pointer by 1.
If $act_i$ is a coding-related action (e.g., the 5th frame in Fig.~\ref{fig:fragment}), it records $f_i$ as the start frame of the coding step to be identified, and analyzes the action and text-line information of $f_i$ and the subsequent frames to identify the coding step.
After outputting a coding step, the approach continues to process the next frame.

\begin{figure}
	\centering
	\includegraphics[width=\linewidth]{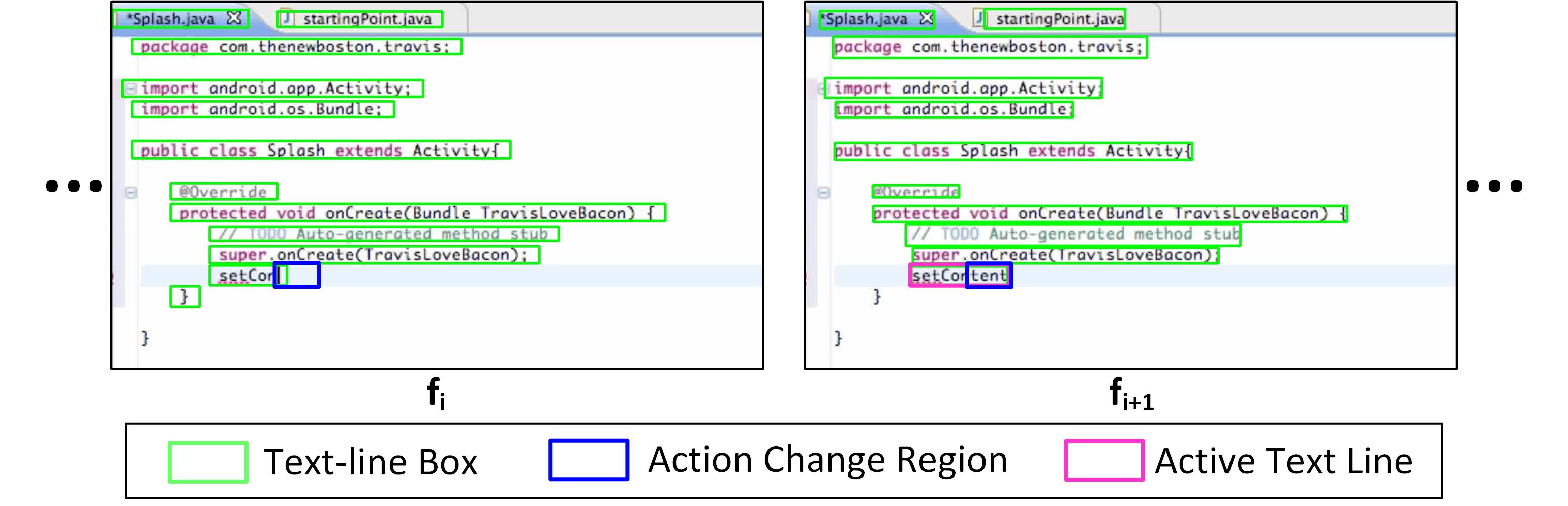}
	\caption{An example of locating active text line(s)}
	\vspace{-1em}
	\label{fig:box}
\end{figure}

As the coding step to be identified is enter-text, delete-text, edit text or select-text, our approach first locates the text line(s) $L_i$ in $f_i$ and $L_{i+1}$ in $f_{i+1}$ affected by the coding-related action $act_i$.
This is determined by overlapping the change regions of $act_i$ with the text-line boxes in $f_i$ (or $f_{i+1}$).
As shown in Fig.~\ref{fig:box}, the area of an action change region and the affected text line(s) may differ greatly, for example, only a few characters are added in a long code line.
However, the action change region and the affected text line(s) should vertically overlap, no matter how different their areas are.
Therefore, our approach identifies a set of vertically continuous text lines (denoted as $L$) with the maximum vertical overlap with the action change region (denoted as $R$).
Multiple text lines are possible as some coding actions may affect a block of text as a whole. 
The vertical overlap $vo$ of $L$ and $R$ is $[R.y_1, R.y_2] \cap [L.y_1, L.y_2] / [R.y_1, R.y_2] \cup [L.y_1, L.y_2]$, where $y_1$ and $y_2$ are the vertical coordinates of the top and bottom line of $L$ or $R$.
If $vo>0.75$, $L$ is identified as active text line(s) for $act_i$.
The threshold 0.75 is determined by examining the correct and wrong overlaps in our dataset.

If $L_{i+1}$ for enter/select-chars includes multiple text lines (e.g., the fourth frame in Fig.~\ref{fig:fragment}), our approach outputs $f_i$ and $f_{i+1}$ as a coding step.
The step type is the corresponding action type.
If $L_{i+1}$ for enter/select-content is empty, $act_i$ is regarded as an invalid action (usually an erroneously recognized coding action).
That is, active text-line analysis helps to filter out erroneous coding actions by ActionNet. 
If $act_i$ is delete-chars and $L_i$ includes multiple text lines, our approach outputs $f_i$ and $f_{i+1}$ as a delete-text step.

If $L_{i+1}$ for a coding action includes only one text line, our approach attempts to aggregate continual coding actions on the same text line (e.g., the 1st and 3rd step in Fig.~\ref{fig:fragment}).
If the action $act_{i+1}$ of $f_{i+1}$ and $f_{i+2}$ is a coding action, the approach identifies active text line(s) $L_{i+1}$ for $act_{i+1}$.
If $L_{i+1}$ contains one text line, the approach determines if $L_i$ and $L_{i+1}$ is a pair of matched text lines between $f_{i+1}$ and $f_{i+2}$.
If that is the case, $act_{i+1}$ and $act_i$ are continual coding actions on the same text line.
Otherwise, the approach computes the vertical overlap of $L_i$ and $L_{i+1}$. 
If the vertical overlap is greater than 0.75, $L_i$ and $L_{i+1}$ are also considered as matched text lines.
This helps to deal with large text edits on the same text line, resulting from text paste or the use of code completion assistant.

The action aggregation stops when a non-coding-related action (except scroll-content) is encountered or $L_i$ and $L_{i+1}$ cannot be matched in neither way.
If a scroll-content action is encountered, our approach examines if $L_i$ has a matched text line $L_{i+1}$ on $f_{i+1}$.
If so, the algorithm continues with $L_{i+1}$ as the active text line for aggregating subsequent coding actions as illustrated in Fig.~\ref{fig:scroll}.
When the aggregation stops, the approach records the latter frame of the last aggregated coding action as the end frame of the coding step.
If active text line in the start frame is empty, the code step is marked as enter-text (the 2nd step in Fig.~\ref{fig:fragment}), otherwise as edit text (the 3rd step in Fig.~\ref{fig:fragment}).
If $L_{i+1}$ is empty (i.e., the entire line is deleted like the 2nd frame in Fig.~\ref{fig:fragment}), the coding step is marked as delete-text.
The algorithm uses the active text line of the last coding action as the text line affected by the coding step.

%% file: Evaluation.tex
\section{Experiment Setup}
\label{sec:experiment}

This section describes our experimental dataset and evaluation metrics for evaluating our workflow-extraction approach.

\subsection{Coding-Step Dataset}
We manually label 41 hours of programming screencasts created by eight different developers, and identify 5,466 line-granularity code-line editing steps with variant durations.
This process costs us about 2 weeks including labeling and checking. 
To the best of our knowledge, our dataset is the first of its kind for evaluating the extraction of code-line editing steps from programming screencasts.

\subsubsection{Data Collection}
\label{sec:datacollection}

We collect programming screencasts from two sources: YouTube and Twitch.
We search YouTube for the programming tutorials on Python and Java (the two most popular programming languages).
We select the returned playlists that demonstrate programming tasks using software development tools, rather than those explaining programming concepts and knowledge on black board or slides.
To test the capability boundary of our approach, we select the playlists using different software development tools (e.g., Eclipse, PyCharm, Notepad and Terminal), recorded with different computer settings (e.g., screen resolution, window theme and text style), and demonstrating diverse development tasks (e.g., programming basics, mobile, web or game development).
In addition, text content on screen has different types of font, size and background.
Finally, we select six playlists (three for Python and three for Java) by six developers, which contain 260 videos with over 31 hours total duration 
(see Table~\ref{tab:dataset}).
The duration of the videos ranges from 4 to 15 minutes (median 8 minutes).

\begin{table}
	\centering
	\caption{Details of our programming screencast dataset}
	\vspace{-0.5em}
	\begin{tabular}{|c|c|c|c|c|c|}
		\hline
		\multicolumn{6}{|c|}{\textbf{Programming videos from YouTube}} \\
		\hline
		\textbf{Language} & \textbf{ID} & \textbf{video} & \textbf{Dur.(h)} & \textbf{Tool} & \textbf{Task} \\
		\hline
		\multirow{3}{*}{Python} & P1 & 52 & 4.98 & \tabincell{c}{Terminal \& \\ Notepad} & \tabincell{c}{Programming \\ basics} \\
		\cline{2-6}
		& P2 & 43 & 3.95 & Pycharm & \tabincell{c}{Programming \\ basics} \\
		\cline{2-6}
		& P3 & 17 & 3.22 & \tabincell{c}{Terminal \& \\ Notepad} & \tabincell{c}{Programming \\ basics} \\
		\hline
		\multirow{3}{*}{Java} & P4 & 99 & 10.08 & Eclipse & \tabincell{c}{Android \\ development}  \\
		\cline{2-6}
		& P5 & 11 & 2.99 & \tabincell{c}{Android \\ studio} & \tabincell{c}{Android \\ development}  \\
		\cline{2-6}
		& P6 & 38 & 5.83 & Eclipse & \tabincell{c}{Web app \\ development}   \\
		\hline
		\multicolumn{6}{|c|}{\textbf{Live coding streams from Twitch}} \\
		\hline
		\multirow{2}{*}{Multiple} & L1 & 5 & 5.00 & \tabincell{c}{Online\\IDE} & \tabincell{c}{Web app \\ development} \\
		\cline{2-6}
		& L2 & 5 & 5.00 &  \tabincell{c}{Sublime \& \\ Shell} & \tabincell{c}{Game \\ development} \\
		\hline
		\textbf{Total} & \textbf{8} & \textbf{270} & \textbf{41.05} & - & - \\
		\hline
	\end{tabular}
	\vspace{-1.5em}
	\label{tab:dataset} \\
	\scriptsize
\end{table}%

We collect 10 live coding sessions on Twitch by the two developers and each session is 1 hour.
The two developers use very different computer settings and tools.
Different from YouTube tutorials which demonstrate a specific task in a video, live coding sessions record the actual, continual development work by developers.
In our collected streams, the two developers use Python, Java and/or Javascript, and occasionally use some shell scripts (e.g., PowerShell).
Live coding developers may switch between programming languages and development tools for different development tasks during a living coding session.

\subsubsection{Manual Labeling}

We decode the collected 270 programming screencasts into 270 sequences of screenshots at the rate of 1 frame per second (fps) by the ffmpeg tool~\cite{ffmpeg}.
We manually label the resulting 41,591 non-identical frames into coding steps, in terms of the start and end frame and the type of each coding step.
The labeling uses the same coding-step identification criteria as described in Section~\ref{sec:codingstepidentification}.
That is, we use human annotators as the most accurate ``computer-vision'' technique to identify ground-truth coding steps for evaluating our computer-vision based approach.
The two authors label the coding steps independently. 
We use the Fleiss' kappa~\cite{fleiss1971measuring} measure to examine the agreement between them. 
The overall kappa value is 0.93, which indicates almost perfect agreement between the labelers.
After completing the labeling process, the two labelers discuss the disagreements to obtain the final labels.

We obtain 5,466 coding steps, with the duration of 7.62 (mean) $\pm$ 10.82 (standard deviation) frames (i.e., seconds at 1 fps).
There are 5,457 non-coding video fragments, with the duration of 16.17$\pm$17.88 frames.
Among the 5,466 coding steps, enter-text, delete-text, edit-text and select-text account for 36.5\%, 0.5\%, 20.9\%, and 42.1\%, respectively.
Delete-text is much less than the other three types because developers either create new programs or modify existing code.
Select text occurs relatively frequently because developers often select code to explain in YouTube video tutorials.

\begin{figure}
	\centering
	\includegraphics[width=\linewidth]{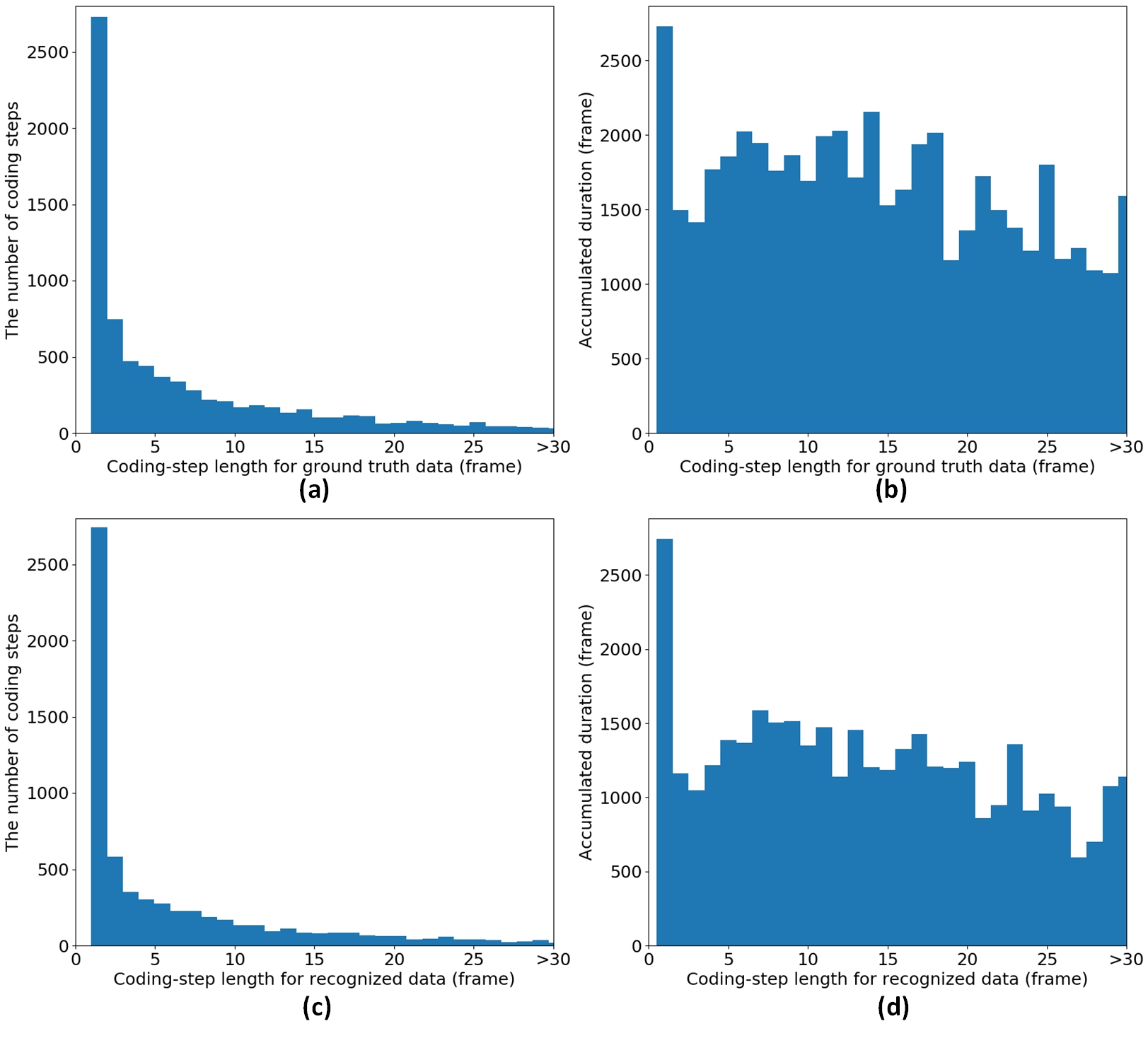}
	\vspace{-2em}
	\caption{Distribution of coding-step length and duration}
	\vspace{-1.5em}
	\label{fig:frag_len_all}
\end{figure}

Fig.~\ref{fig:frag_len_all}(a) shows the distribution of coding-step length. 
The shortest coding step contains two frames.
Coding steps with 5 or less frames account for about 61.7\% of the 5,466 ground-truth coding steps.
There are many short coding steps because developers often make small modifications to existing lines of text, or make atomic text-block modifications.
However, as shown in Fig.~\ref{fig:frag_len_all}(b), the accumulated duration of different lengths of coding steps does not differ as much as the frequency of different code-step lengths.
The accumulated duration of coding steps with 5 or less frames account for only 18.6\% of the total duration of the 5,466 coding steps.

\subsection{Evaluation Metrics}

We compute Precision, Recall and F1-score of the identified coding steps in a programming screencast against the ground-truth coding steps.
Precision is the portion of coding steps that are correctly recognized among all the identified coding steps.
Recall is the portion of the ground-truth coding steps that are correctly recognized.
F1-score conveys the balance between Precision and Recall which is computed by $2 \times (Precision  \times Recall) / (Precision + Recall)$.

To determine the correctness of an identified coding step, we need to first determine the accuracy of the boundary (i.e., start and end frame) of the identified coding step against the ground-truth coding step.
To that end, we compute Intersection over Union (IoU) of the two video fragments.
Let $r$ be the coding-step fragment identified by our method, and $gt$ be the ground-truth fragment.
IoU is computed by $r \cap gt/r \cup gt$, i.e., the number of common frames between the two fragments over the total number of frames in the set of two fragments.
If $r$ and $gt$ have no overlapping, their IoU is 0.

Given the sequence of identified coding-step fragments $R = \{r_1, r_2, \cdots, r_m\}$ and the sequence of ground-truth fragments $G = \{gt_1, gt_2, \cdots, gt_n\}$ of a programming screencast, we compute the longest common subsequence between $R$ and $G$, with the goal of maximizing the IoU of the matched $r_i$ and $gt_j$.
When a $r_i$ overlaps multiple ground-truth fragments, we select the $gt_j$ with the highest IoU with $r_i$.
When a $r_i$ is matched with a $gt_j$, the matching continues from the frame next to the end frame of $r_i$.
We allow multiple $r_i$ to be matched to different fragments of one $gt_j$.
This could happen when a long ground-truth fragment has been recognized as several consecutive short fragments (i.e., over-segmentation).

Given a pair of matched coding-step fragments $r_i$ and $gt_j$, if their IoU is above a threshold, we consider $r_i$ as a correctly identified coding-step fragment.
The higher the IoU threshold is, the more accurate the boundary of the identified coding-step fragment is, but the less number of fragments can be qualified as correct.
If the step type of a correctly identified $r_i$ matches the step type of $gt_j$, the coding step of $r_i$ is regarded as a correctly identified coding step.

IoU has been widely used for object detection in computer vision. 
As described in \cite{kisantal2019augmentation, yan2019iou}, even the best matching anchor box of a small object has a low IoU value typically.
Similarly, IoU is sensitive to short fragment length.
For example, assume $gt$ has 4 frames and $r$ has 3 frames.
The highest IoU is only 0.75, when all three frames in $r$ overlap $gt$.
When IoU threshold is above 0.75, the identified fragment $r$, albeit high quality, will be regarded as incorrect.
Therefore, we compute time offset as a complementary metric to understand the mismatch between the two overlapping fragments.
When IoU $>$ 0, the time offset is computed by $min(|F_r.start-F_{gt}.start|, |F_r.end-F_{gt}.start|, |F_r.start-F_{gt}.end|, |F_r.end-F_{gt}.end|)$.
Time offset represents the least effort required to navigate from the start or end frame of the identified coding step to the start or end frame of the overlapping ground-truth step.
As shown in Fig.~\ref{fig:frag_len_all}, the majority of coding steps are short.
Therefore, it would not be difficult to identify the complete coding step after locating its start or end frame.
Given a pair of matched $r_i$ and $gt_j$, if their time offset is below a threshold, we consider $r_i$ as a correctly identified coding-step fragment.

\section{Experiment Results and Findings}
\label{sec:result}

We conduct extensive experiments on our coding-step dataset to investigate the following four research questions: 

\begin{itemize}
	\item \textbf{RQ1.} How do the coding-step trailers extracted by our approach compare with the ground truth trailers?
	
	\item \textbf{RQ2.} How well does our approach perform under different IoU and time-offset threshold settings?
	
	\item \textbf{RQ3.} How well does our approach perform for different developers, development environments, programming languages, and programming tasks? 
	
	\item \textbf{RQ4.} How well do developers rate the quality of the identified coding steps?	
	
\end{itemize}

\begin{table}
	\centering
	\caption{Performance at different matching thresholds}
	\vspace{-0.5em}
	\begin{tabular}{|c|c|c|c|c|c|c|c|}
		\hline
		\multicolumn{4}{|c|}{\textbf{IOU thresholds}} & \multicolumn{4}{|c|}{\textbf{Time-offset thresholds}} \\
		\hline
		\textbf{IoU} & \textbf{Prec} & \textbf{Reca} & \textbf{F1} & \textbf{TO} & \textbf{Prec} & \textbf{Reca} & \textbf{F1} \\
		\hline
		$>$0 & 0.891 & 0.942 & 0.909 & =0 & 0.838 & 0.824 & 0.820 \\
		\hline
		$>$0.3 & 0.772 & 0.738 & 0.744 & $\leq$1 & 0.869 & 0.864 & 0.856 \\
		\hline
		$>$0.5 & 0.703 & 0.671 & 0.677 & $\leq$3 & 0.882 & 0.891 & 0.876 \\
		\hline
		$>$0.7 & 0.612 & 0.586 & 0.590 & $\leq$5 & 0.890 & 0.911 & 0.891 \\
		\hline
		$>$0.9 & 0.548 & 0.525 & 0.529 & $\leq$7 & 0.892 & 0.926 & 0.900 \\
		\hline
		=1.0  & 0.520 & 0.498 & 0.501 & $\leq$9 & 0.894 & 0.937 & 0.907 \\
		\hline
	\end{tabular}
	\vspace{-1em}
	\label{tab:iou} \\
	\scriptsize
\end{table}%

\subsection{Comparison of Extracted and Ground-Truth Trailers (RQ1)}
\label{sec:rq1}
\subsubsection{Motivation}
A coding-step trailer provides a concise overview of coding steps in a programming screencast.
Each screencast has an extracted trailer and a ground-truth trailer.
We want to investigate how the identified coding steps compare with the ground-truth coding steps at the trailer level.


\subsubsection{Method}
We make three types of comparisons between the extracted trailers $T_i$ and the ground-truth trailers $T_{gt}$.
First, we compare the distribution of coding-step length and total duration in $T_i$ versus $T_{gt}$.
Second, we compare the frame coverage of a set of trailers ($T_i$ or $T_{gt}$) over the original programming screencasts.
Third, we compute three percentages: the overall IoU of $T_i$ and $T_{gt}$, the percentage of the frames only in $T_i$ (false positive), and the percentage of the frames only in $T_{gt}$ (false negative). 

\subsubsection{Results}
Our approach identifies 4,648 coding steps in the 270 programming screencasts.
Fig.~\ref{fig:frag_len_all}(c) and Fig.~\ref{fig:frag_len_all}(d) show the frequency and total duration of different lengths of the identified coding steps, respectively.
Compared with those of the ground-truth coding steps in Fig.~\ref{fig:frag_len_all}(a) and Fig.~\ref{fig:frag_len_all}(b), we observe similar overall distribution.
About 66.0\% of the identified coding steps have 5 or less frames.
This percentage is relatively higher than the percentage of the ground-truth counter part (61.7\%).
Furthermore, the ground-truth trailers have longer coding steps than the extracted trailers, and relatively longer total duration.
This is because our approach may identify several shorter coding steps for a long coding step.
However, the overall frame coverage does not differ significantly, with 26.9\% for the extracted trailers and 28.1\% for the ground-truth trailers.
The overall IoU of the extract trailers and the ground-truth trailers is 0.661, only 22.4\% of the frames in the extracted trailers are not in the ground-truth trailers, and only 21.3\% of the frames in the ground-truth trailers are not covered in the extracted trailers.

\vspace{1mm}
\noindent\fbox{\begin{minipage}{8.4cm} \emph{The coding steps identified by our approach have similar length and duration distributions to the ground-truth steps. They also have similar overall coverage of the programming screencasts and low false-positive and false-negative frames.} \end{minipage}}\\

\subsection{Performance Under Different Matching Thresholds (RQ2)}
\label{sec:rq2}

\subsubsection{Motivation}
The correctness of coding steps depends on the IoU (or time-offset) threshold.
The strictness of the threshold affects the boundary accuracy and the number of coding-step fragments that can be treated as correct, which in turn affects the performance of coding step extraction.
In this RQ, we investigate the impact of the IoU and time-offset threshold on the extraction of coding steps.

\subsubsection{Method}
We experiment six IoU thresholds \{0, 0.3, 0.5, 0.7, 0.9, 1.0\}.
IoU $>$ 0 means the identified coding step and the ground-truth step have at least one overlapping frame, and IoU = 1.0 means the identified step and the ground-truth step have identical frames from start to end.
We experiment six time-offset thresholds \{0, 1, 3, 5, 7, 9\} (frame).
time-offset = 0 means the identified coding step and the ground-truth step have at least one matched boundary, and time-offset $\leq$ 9 means the closest boundaries of the identified step and the ground-truth step have at most 9 frames gap.
At each threshold, we obtain a set of correctly identified coding steps.
We compute precision, recall and F1-score of these identified steps.



\subsubsection{Results}

Table \ref{tab:iou} shows the overall evaluation metrics of all 4,648 identified coding steps at different IoU thresholds.
At the strictest threshold IoU = 1.0, our approach still achieves 0.5 F1-score, which means about half of the identified coding steps match perfectly with the corresponding ground-truth steps.
As the IoU threshold decreases, the boundary accuracy criteria is loosen, and more and more identified coding steps are qualified as correct.
Consequently, the F1 increases gradually from 0.5 for IoU = 1.0 to 0.74 for IoU $>$ 0.3.
At the lowest threshold IoU $>$ 0, the F1-score reaches the upper bound 0.9.
That is, about 90\% of the identified coding steps have at least one overlapping frame with the ground-truth steps.

\begin{figure}
	\centering
	\includegraphics[width=\linewidth]{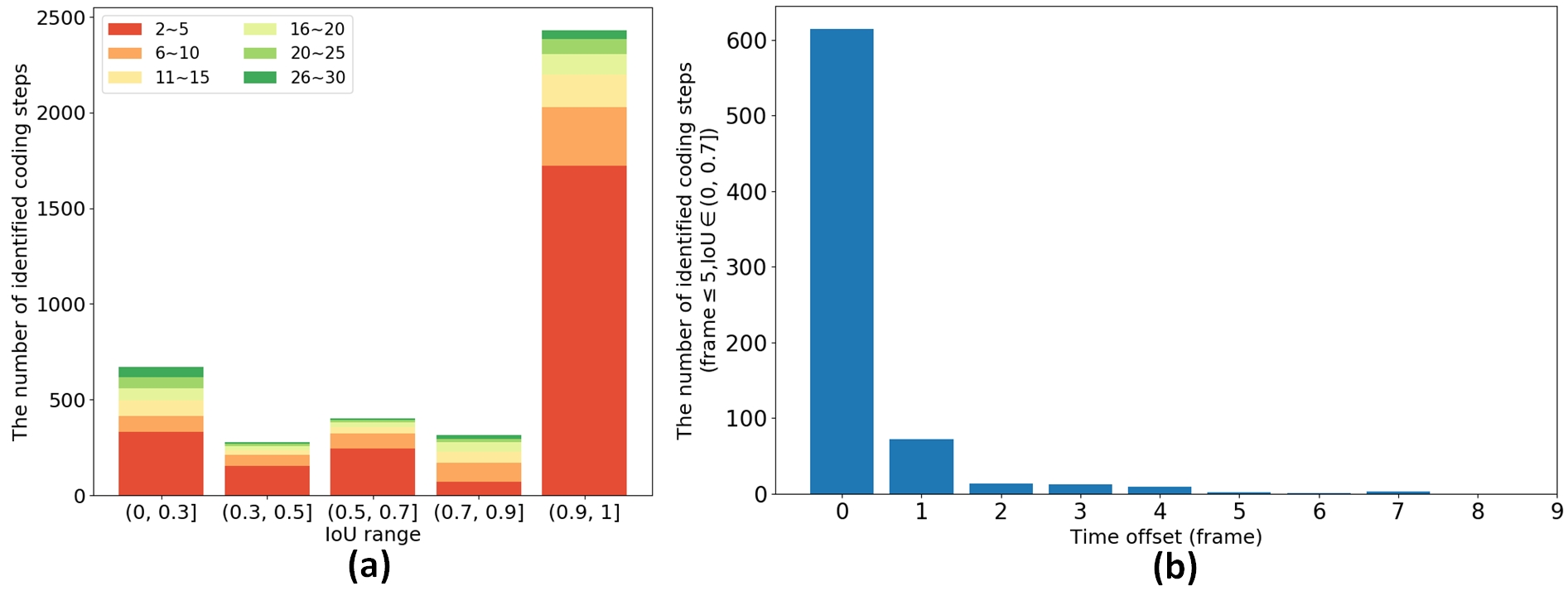}
	\caption{Coding-step distribution at IoU and time-offset  thresholds}
	\vspace{-1.5em}
	\label{fig:codingstep}
\end{figure}

Fig.~\ref{fig:codingstep}(a) is the distribution of different lengths of identified coding steps at different IoU threshold ranges.
About 71.0\% of coding steps with 5 or less frames (red bar) have IoU$\in (0.7, 1.0]$ with the ground-truth steps.
Only a small number of coding steps with 5 or less frames have IoU$\in (0.7, 0.9]$, but about 13.2\% of coding steps with 5 or less frames have IoU$\in (0, 0.3]$.
This is because it is hard for short coding steps to achieve high IoU when they do not completely match the ground-truth steps.
In contrast, about 60.4\% of coding steps with more than 5 frames have IoU$\in (0.7, 1.0]$, and the rest of such long coding steps are roughly evenly distributed in different IoU ranges.
This is because IoU metric is more robust when the length of coding steps increases.

For time-offset based performance metrics, as shown in Table~\ref{tab:iou}, we can see that the upper bound of time-offset based metrics at time-offset $\leq$ 9 is almost the same as those at IoU $>$ 0.
However, when the time-offset threshold becomes stricter, the metrics decreases much slower, compared with the metric decrease as IoU increases.
At the strictest time-offset = 0, the F1 is 0.82, which means about 82\% of the identified coding steps have at least one boundary matching the boundary of the corresponding ground-truth steps.

Furthermore, time-offset threshold at 5 or above improves the performance very marginally.
This suggests that the boundary of correctly identified coding steps is usually very close (1-4 frame gap) to that of the ground-truth steps.
Fig.~\ref{fig:codingstep}(b) shows the time-offset distribution of the identified coding steps with 5 or less frames and having IoU$\in (0, 0.7]$ with the ground-truth steps.
94.2\% of these coding steps have time-offset = 0 or 1.
In fact, this is the main reason why time-offset based performance is much better than IoU based.

\begin{table}
	\centering
	\caption{Performance on different playlists}
	\vspace{-0.5em}
	\begin{tabular}{|c|c|c|c|c|c|c|c|}
		\hline
		\multicolumn{8}{|c|}{\textbf{Programming videos from YouTube}} \\
		\hline
		\multicolumn{2}{|c|}{\textbf{}} & \multicolumn{3}{c}{\textbf{IoU $>$ 0.7}} & \multicolumn{3}{|c|}{\textbf{Time-offset $\leq$ 3}} \\
		\hline
		\textbf{Language} & \textbf{Idx} & \textbf{Prec} & \textbf{Reca} & \textbf{F1} & \textbf{Prec} & \textbf{Reca} & \textbf{F1} \\
		\hline
		\multirow{4}{*}{Python} & P1 & 0.703 & 0.696 & 0.691 & 0.936 & 0.932 & 0.923 \\
		\cline{2-8}
		& P2 & 0.645 & 0.553 & 0.585 & 0.917 & 0.838 & 0.863  \\
		\cline{2-8}
		& P3 & 0.705 & 0.644 & 0.670 & 0.905 & 0.902 & 0.901  \\
		\cline{2-8}
		& Ave. & 0.684 & 0.631 & 0.648 & 0.919 & 0.891 & 0.896 \\
		\hline
		\multirow{4}{*}{Java} & P4 &0.535 & 0.526 & 0.522 & 0.883 & 0.905 & 0.880  \\
		\cline{2-8}
		& P5 & 0.505 & 0.512 & 0.495 & 0.811 & 0.859 & 0.820  \\
		\cline{2-8}
		& P6 & 0.578 & 0.588 & 0.579 & 0.842 & 0.910 & 0.871  \\
		\cline{2-8}
		& Ave. & 0.539 & 0.542 & 0.532 & 0.882 & 0.891 & 0.876 \\
		\hline
		\multicolumn{8}{|c|}{\textbf{Live coding streams from Twitch}}  \\
		\hline
		\multirow{3}{*}{Multiple} & L1 & 0.601 & 0.649 & 0.615 & 0.938 & 0.833 & 0.882  \\
		\cline{2-8}
		& L2 & 0.625 & 0.526 & 0.571 & 0.923 & 0.800 & 0.857  \\
		\cline{2-8}
		& Ave. & 0.613 & 0.587 & 0.593 & 0.930 & 0.816 & 0.869 \\
		\hline
	\end{tabular}
	\vspace{-1em}
	\label{tab:playlist} \\
	\scriptsize 
\end{table}%

\vspace{1mm}
\noindent\fbox{\begin{minipage}{8.4cm} \emph{Our approach can accurately extract coding steps, even at the strictest matching criteria. A large portion of short coding steps have low IoU with the ground-truth steps, but they have only 0-1 frame time offset.} \end{minipage}}\\

\subsection{Performance On Diverse Programming Screencasts (RQ3)}
\subsubsection{Motivation}
Developers use different programming languages and tools in their work, and their computer settings (e.g., screen resolution, window theme and text style) also vary.
There are also many different types of programming tasks.
This RQ aims to investigate the generality of our approach in such diverse settings.

\subsubsection{Method}
Our coding-step dataset consists of 8 sets of programming screencasts, created by 8 different developers with diverse computers, tools, programming languages and task settings (see Table~\ref{tab:dataset}).
We first analyze the performance variations across 270 programming screencasts at different IoU and time-offset thresholds.
Then, we zoom into the performance of each set of programming screencasts at the IoU $>$ 0.7 or time-offset $\leq$ 3.

\subsubsection{Results}

\begin{figure}
	\centering
	\includegraphics[width=\linewidth]{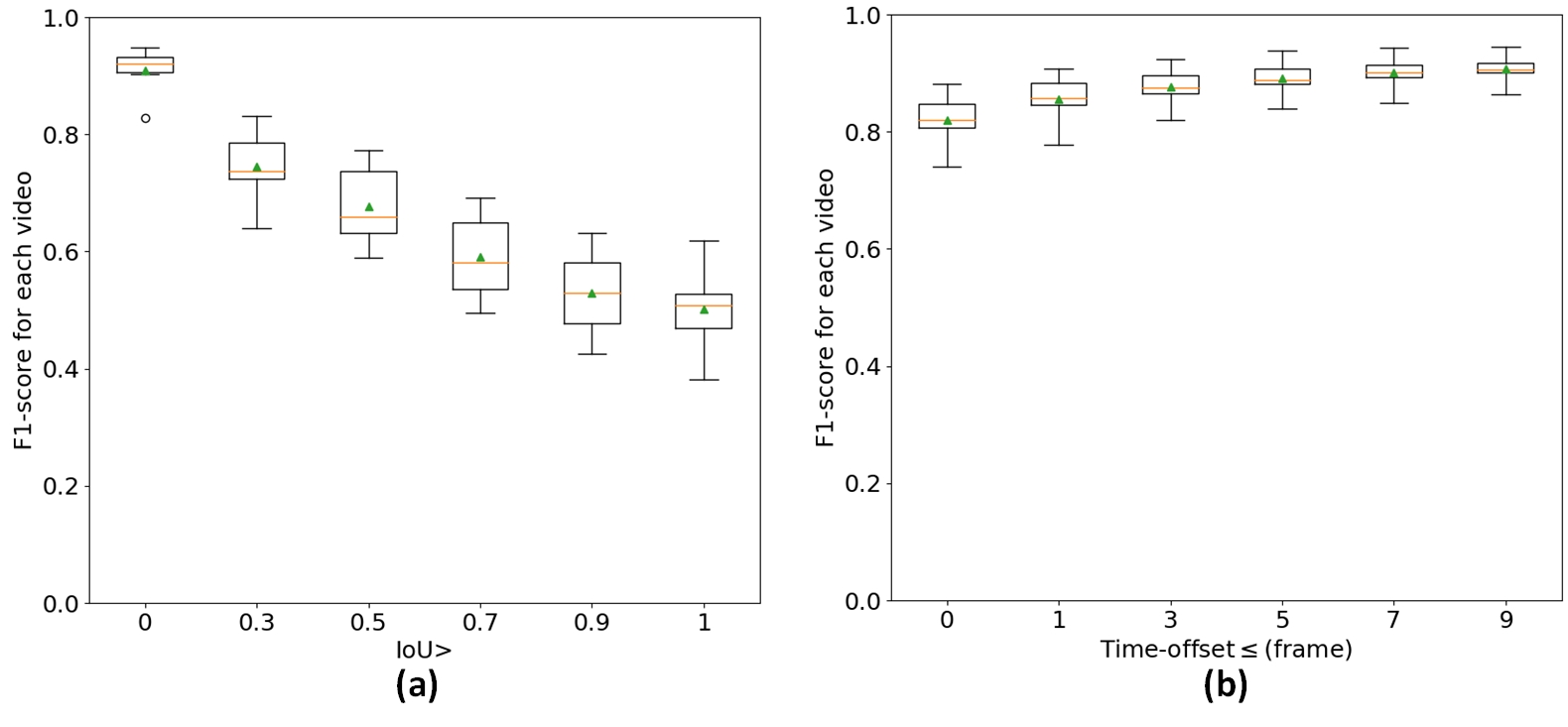}
	\vspace{-2em}
	\caption{Per-video F1 distribution at different IoU and time-offset thresholds}
	\label{fig:f1distribution}
\end{figure}

Fig.~\ref{fig:f1distribution}(a) and Fig.~\ref{fig:f1distribution}(b) show the distribution of the F1-score of 270 programming screencasts at different IoU and time-offset thresholds.
The interquartile range of the box plots is mostly 0.1 between the 25th percentile and the 75th percentile for IoU-based evaluation, and less than 0.05 for time-offset based evaluation.
Some screencasts are more challenging than others, which results in about 0.2 difference between the best and the worst F1-score for IoU, and about 0.1 F1-score difference for time-offset.
Time-offset based evaluation is more stable than IoU based evaluation, due to the time-offset's tolerance of small mismatches.

Table \ref{tab:playlist} shows the metrics on the eight sets of programming screencasts.
The performance rankings of these eight sets are consistent across IoU$>$0.7 and time-offset$\leq$3 threshold.
This indicates the consistency between the two thresholds, even though they represent different criteria for determining the correctness of the identified coding steps.
P1 has the best performance, and P5 has the worst performance.
Compared with the overall average F1-score (0.59 for IoU$>$0.7 and 0.876 for time-offset$\leq$3), only P1 and P5 have larger variation ($\pm$0.1 in F1-score for IoU$>$0.7 and $\pm$0.05 in F1-score for time-offset$\leq$3).
The 3 Python playlists have better performance, the 3 Java playlists have worse performance, and the 2 live coding sets are in between.

\begin{figure}
	\centering
	\includegraphics[width=\linewidth]{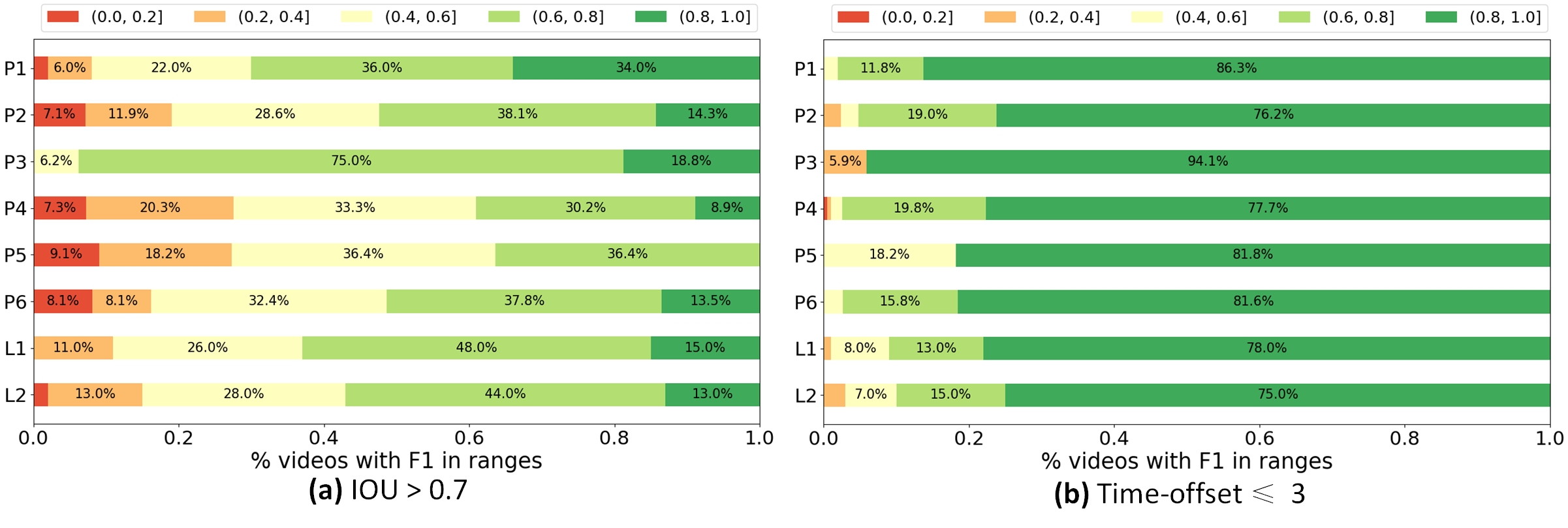}
	\vspace{-2em}
	\caption{Coding-step distribution at different IoU and time-offset thresholds}
	\vspace{-1em}
	\label{fig:playlist}
\end{figure}

Fig.~\ref{fig:playlist} shows the percentage of videos in each set of programming screencasts whose F1-scores fall into different ranges.
We show percentage because different sets have different numbers of screencasts.
For 6 sets of programming screencasts, more than half of the videos have F1-score $>$ 0.6 (light green and green bars) for IoU-based evaluation.
For all 8 sets of programming screencasts, more than 81.3\% of the videos have F1-score $>$ 0.8 (green bars) for time-offset based evaluation.
The videos with low F1-scores ((0.0,0.2] in red or (0.2,0.4] in orange) usually account for a small portion of a set of programming screencasts.
This is especially the case for time-offset based performance.

We examine the videos with low F1-scores to understand what causes the low accuracy.
We find that the challenging screencasts usually involve complex human computer interactions.
For example, P6 demonstrates Web app development, in which the developer frequently switch between web design tool, command-line terminal and web browser to edit, deploy and view the web page back and forth. Sometimes, different windows may even overlap.
As another example, Java IDEs (e.g., Eclipse) often provide sophisticated pop-ups.
These pop-ups blur the coding steps with very dynamically changed pop-up contents.
In contrast, Python developers often use text editors (e.g., Notepad) where coding steps are clear to identify.
Complex human computer interaction poses a significant challenge for ActionNet to accurately recognize primitive HCI actions of interests.
Some meaningless HCI actions may also confuse our approach.
For example, some developers like to randomly select some code fragments, switch windows or scroll content while thinking about code features or errors, which result in many meaningless actions.
The erroneously recognized or meaningless HCI actions negatively affect the downstream inference of coding steps. 

\vspace{1mm}
\noindent\fbox{\begin{minipage}{8.4cm} \emph{Our approach performs stably across diverse programming screencasts. It can accurately identify coding steps in over 81\% of the programming screencasts in our dataset by time-offset based evaluation. Improving the ActionNet's recognition capability for complex human computer interaction could further improve the performance.} \end{minipage}}\\

\subsection{Coding Step Quality by Human Evaluation (RQ4)}

\begin{figure}
	\centering
	\includegraphics[width=\linewidth]{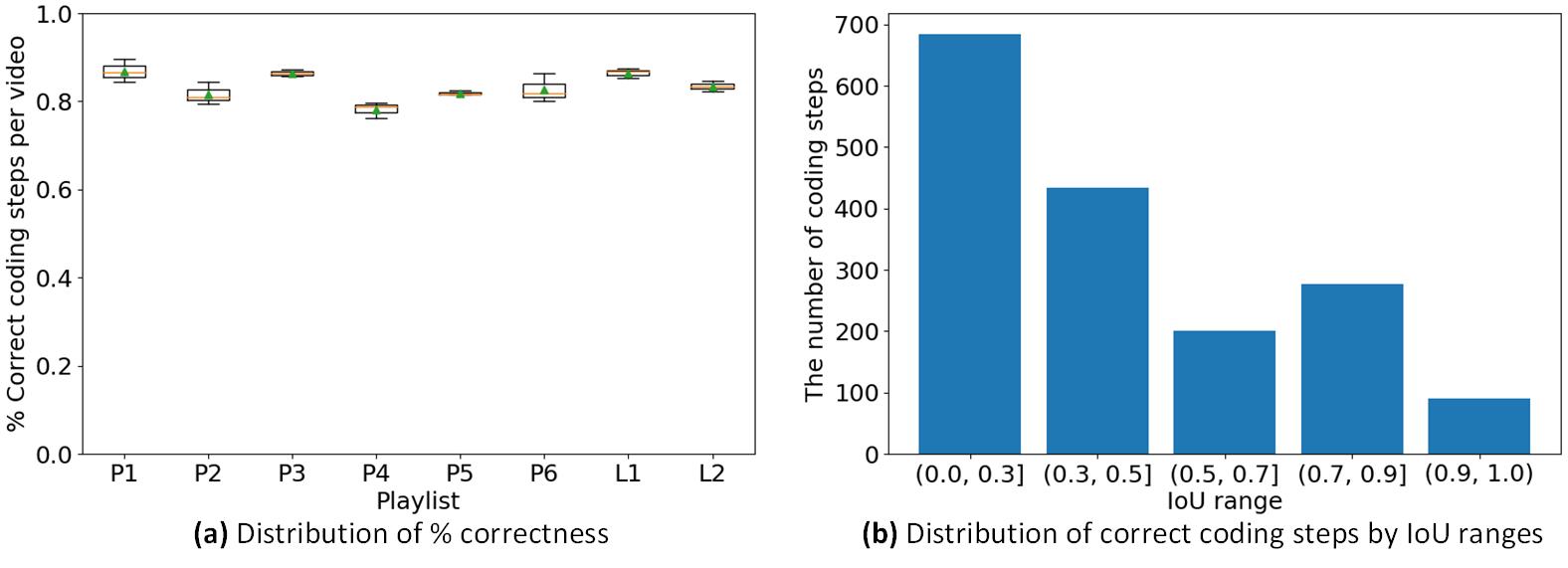}
	\vspace{-2em}
	\caption{Correctness ratings by human evaluation}
	\vspace{-1em}
	\label{fig:usefulness}
\end{figure}

\subsubsection{Motivation}
The identified coding steps may not perfectly match the ground-truth coding steps at both boundaries.
The evaluation against these ground-truth coding steps, no matter IoU or time-offset based, mechanically judge the correctness of the identified coding steps by an overlapping threshold.
This RQ aims to investigate how developers judge and understand the correctness of not-perfectly-aligned coding steps.

\subsubsection{Method}
We exclude the identified coding steps with IoU = 1.0 as they perfectly match the ground-truth coding steps labeled by the two authors.
For the 270 programming screencasts in our dataset, we keep the 2,605 not-perfectly-aligned coding steps (i.e., IoU$\in (0, 1.0)$).
We recruit 6 developers who have at least 3 years programming experience on Python and Java from a top technology company.
The way SeeHow works is kept confidential before all of them finish the experiment.
We ask them to watch a programming screencast in our dataset and then review the non-perfectly-aligned coding steps for this screencast.
The developers can contrast these coding steps against the original programming screencast and/or the ground-truth coding steps.
They give a binary rating for each non-perfectly-aligned coding steps: correct or incorrect.
Correct rating means the developers believe the inaccurate boundaries of the identified coding steps do not affect their correct understanding of the identified steps.
The 6 developers examine all 270 programming screencasts in our dataset, and perform the correctness annotation independently.
We compute Fleiss's Kappa~\cite{fleiss1971measuring} to evaluate the inter-rater agreement, and use the majority of three ratings for a step as the final rating.

\subsubsection{Results}

The 6 developers have substantial agreement (Fleiss' kappa = 0.83) for marking the correctness of the not-perfectly-aligned coding steps.
Fig.~\ref{fig:usefulness}(a) shows the distribution of the percentage of correct-rated coding steps per video in the 8 sets of programming screencasts.
The overall average correctness percentage is 83.6\%.
This correctness percentage is for not-perfectly-aligned coding steps, but it is close to the overall F1-score 79.3\% at Io$>$0.2 and the overall F1 85.6\% at time-offset$\leq$1.
The correctness percentage has small variations within and across the 8 sets of programming screencasts.
The correctness percentage of Python programming tutorials and live coding streams are slightly higher than that of Java programming tutorials.
These correctness rating results are consistent with those by IoU and time-offset based evaluation.

Fig.~\ref{fig:usefulness}(b) shows the distribution of the correct-rated coding steps in different IoU ranges with the ground-truth coding steps.
We see that correct-rated coding steps are scattered in all IoU ranges.
About 700 correct-rated coding steps have $\leq$ 0.3 IoU with the ground-truth coding steps.  
Those coding steps are regarded as incorrect steps at IoU$>$0.3.
But they are rated as correct when reviewed by the 6 developers.
As discussed in Section~\ref{sec:rq2}, a large portion of identified coding steps with low IoU have a small boundary gap (0-3 frames) from the ground-truth steps.
When watching a coding step in the programming screencast, such a small gap does not affect the understanding of the identified coding steps. 

\vspace{1mm}
\noindent\fbox{\begin{minipage}{8.4cm} \emph{The developers' correctness ratings of the identified coding steps are similar to time-offset base evaluation. They can understand the identified steps in face of small boundary inaccuracy or over-segmentation.} \end{minipage}}\\

\subsection{Use case scenarios}
In order to explore the potential scenarios to use our tool in practical environment, an interview is conducted with the 6 developers. 
We focus on one question that how can SeeHow help their daily work.
The answers are summarized as two points: software asset management and workflow logging.

\subsubsection{Software asset management}
Programming screencasts are taken as a kind of software asset~\cite{zabardast2021asset} because of the code-intensive characteristic of such media.
However, programming screencasts on video platforms (e.g., YouTube) are analogous to GitHub repository without version control, which is hard to maintain.
Our tool plays the role of software asset management and enables a lot of downstream tasks such as video summary and video search.
For example, it will save developers about 70\% of time if we only keep the workflow-related fragments of programming screencast in our dataset.
In addition, developers can jump to a specific workflow step via keyword search.

\subsubsection{Workflow logging}
IDEs can record developers' programming behaviours such as typing and deleting text.
However, the recorded logs are limited to single application and cross-application logging is a big challenge.
For example, the developer copy a code line from website and paste it to IDE.
Our tool is computer vision based and non-intrusive, which means it works on various text editing applications.
It can log the text editing behaviours and make developers' workflow traceable.

SeeHow is the first attempt to these promising applications and further efforts are required to realize the value from it.

%% file: RelatedWork.tex
\section{Related Work}
\label{sec:relatedwork}

VT-Revolution~\cite{bao2018vt} develops a novel way for making interactive programming screencasts.
During screen recording, it relies on software instrumentation to record code-line editing steps.
The recorded coding steps allows developers to scan, search and navigate programming screencast just like dealing with textual tutorials.
Although the enhanced interaction with programing screencasts is intriguing, its reliance on software instrumentation is a major barrier for the wide adoption of VT-Revolution.
The authors of VT-Revolution envision to make regularly-recorded programming screencasts interactive through computer-vision based workflow extraction.
Our work is a step towards this vision, and our output coding steps can be seamlessly integrated into VT-Revolution.

CodeTube~\cite{ponzanelli2016codetube, ponzanelli2016too} and CodeMotion~\cite{khandwala2018codemotion} also support this vision.
CodeTube extracts code-like content from screenshots and splits a long video into fragments based on the content similarity of adjacent screenshots.
The fragments correspond to coarse-grained programming activities (e.g., all changes in a file) and do not have explicit notion of coding steps.
CodeMotion extracts code content in a similar way to CodeTube, but it attempts to infer more fine-grained coding steps from content changes only.
As CodeMotion is action agnostic, its action inference is unreliable in face of complex window interactions, such as pop-up assisted code editing and content scrolling.
In contrast, our approach is action aware and infers code-line editing steps based on both coding-related actions and text edits.
Our approach can be integrated into CodeTube to enhance coding step search, such as finding when an API call is first introduced or when a parameter of this API call is changed.
Such fine-grained video search is impossible with the current CodeTube's coarse-grained video fragments.

Our method is related to action detection in natural scene~\cite{xu2017r, zhao2017temporal, lin2017single, caba2015activitynet, hou2017tube}, which aims to detect the start and end time of action instances in untrimmed videos.
Bergh et.al \cite{bergh2020curated} creates a dataset comprised of 111,229 screenshots from Java and Python tutorials.
Their goal is to classify four categories of code images: Visible Typeset Code, Partially Visible Typeset Code, Handwritten Code, and No Code.
This is different from our work in two aspects: our tool processes screencasts (i.e., a sequence of screenshots), and our goal is to segment video frames into coding steps. 
ActionNet~\cite{zhao2019actionnet} is a neural model for recognizing primitive HCI actions between two consecutive frames in screencasts, and it does not extract action-related content.
Our approach entends the ActionNet's output and exploits both primitive HCI actions and coding-action-related text edits to infer coding-step fragments, types and affected contents.

Computer-vision techniques have been applied to GUI data, such as GUI widget detection~\cite{chen2020object}, GUI testing~\cite{white2019improving, degott2019learning, bernal2020translating, yandrapally2020near}, GUI code generation~\cite{chen2018ui}, GUI visual defect detection~\cite{zhao2020seenomaly, wu2020predicting, liu2020owl, moran2018automated}, GUI evolution analysis~\cite{moran2018detecting} and GUI content prediction~\cite{ott2018deep, ott2018learning}.
Our work adds to this picture a new way to extract coding steps from screencasts.
In terms of information granularity, SeeHow fills the gap between ActionNet~\cite{zhao2019actionnet} and CodeTube~\cite{ponzanelli2016codetube, ponzanelli2016too}.

\section{Threats to Validity}
One threat to internal validity is the labeling errors of our screencast dataset.
To minimize the labeling errors, two authors label the data independently and reach the almost-perfect agreement.
This high agreement benefits from the clear definition of code-line editing steps the developers are familiar with.
Disagreements mainly come from judging whether some actions (e.g., typing a class name to search code) are coding related.
Another internal threat is the human bias for evaluating non-perfectly aligned coding steps in RQ4.
None of the annotators are involved in our work and they have no knowledge about our goal and approach design.
Furthermore, the ratings by the 6 annotators have high agreements.
Threats to external validity include the representativeness of our screencast data and the choice of text recognition tools.
Our experiments intentionally include very diverse screencasts in terms of developers, programming languages,  tasks and tools, and computer settings.
We carefully chose text recognition tools which are appropriate for GUI texts~\cite{chen2020object}.
These tools may have errors at char level which will not affect the coding step identification at line level.

\section{Conclusion and future work}
\label{sec:conclusion}

This paper presents the first non-intrusive, computer-vision based approach for extracting code-line editing steps from programming screencasts, without the need of software instrumentation.
The innovation lies in action-aware text extraction and coding step identification.
The evaluation on 41 hours diverse programming screencasts shows that half of the coding steps identified by our approach match perfectly with the ground-truth steps, and most of the not-perfectly-matched steps have small (0-3) frame offset at the start or the end.
The 6 developers judge 83\% of the identified coding steps as correct, with substantial inter-rater agreement.
The output coding steps by our approach lower the bar to make interactive programming screencasts which currently has to rely on software instrumentation, and can also be integrated into programming video search engine which currently supports only action-agnostic, coarse-grained content search. 
We will investigate advanced workflow analysis enabled by non-intrusive workflow extraction from screencasts, such as workflow pattern mining and workflow error identification, to support best programming practices.

